# The “Easy Trap”: Why LLMs Underestimate Misconception-Driven Difficulty

Amanda La Hadi
Monash University, Indonesia
amanda.lahadi@monash.edu

Muhammad Johan Alibasa
Monash University, Indonesia
johan.alibasa@monash.edu

Guanliang Chen
Monash University, Australia
guanliang.chen@monash.edu

A. Taufiq Asyhari
Monash University, Indonesia
taufiq.asyhari@monash.edu

## ABSTRACT

Large language models (LLMs) are increasingly used for estimating item difficulty in educational assessment. However, it remains unclear whether such estimates reflect how learners actually experience difficulty. This study investigates the alignment between LLM-generated difficulty ratings and empirical student performance on basic mathematics tasks. Four widely used LLM-based systems generated difficulty ratings (1–100 scale) for 32 arithmetic items across multiple runs (N = 640 ratings). These were compared with empirical difficulty derived from responses of 770 Indonesian undergraduates using Classical Test Theory (CTT) and Item Response Theory (2PL). Results show moderate rank correlations (Spearman’s $\rho = 0.52 - 0.7$), indicating that LLMs capture coarse ordering of item difficulty. However, substantial and systematic misalignment emerges in fraction items. Several items consistently rated as “easy” by LLMs were among the most difficult for students (e.g., only 34.16% correct for $100 \div \frac{1}{2}$). We argue that LLMs approximate curricular difficulty—what should be easy based on instructional sequencing—rather than cognitive difficulty driven by learner misconceptions. This leads to systematic underestimation of misconception-driven items, a phenomenon we term the “Easy Trap.” These findings highlight a critical limitation of LLM-based difficulty estimation and suggest that relying on such estimates without empirical grounding may introduce bias in assessment design and adaptive systems.

## 1. INTRODUCTION

Large language models (LLMs) are rapidly being integrated into educational practice, extending far beyond basic search or item drafting to functions that directly affect assessment—providing feedback, assisting with grading for multiple-choice and open-ended responses, recommending remedial content, generating curriculum-aligned items, and even estimating item difficulty from text alone [4,8,19,27,29,35,38]. While these capabilities promise substantial gains in efficiency, they also introduce crucial measurement risks: LLM-generated judgments often reflect curricular expectations rather than empirical patterns of student performance, potentially leading to mis-calibrated difficulty labels when real learners diverge from what the curriculum prescribes [1,23].

This risk is becoming salient for so-called “basic” mathematics. Foundational number concepts—arithmetic with whole numbers, integers, fractions, decimals, and exponents—appear early in schooling yet continue to challenge many undergraduates, with well-documented downstream effects on later success in STEM disciplines [5,16,21,33,36]. Among these domains, fractions are repeatedly identified as a bottleneck skill and a strong predictor of attainment in algebra, statistics, and calculus [5,14,21,30]. Formal instruction in fractions begins in Grade 3, but persistent misconceptions—e.g., whole-number bias, treating numerators and denominators independently, or assuming “division makes smaller”—are widely reported across contexts and often persist into higher education [20,22]. When LLMs classify such fraction items as “easy” on the grounds that they are introduced early in curricula, their predictions may systematically underestimate actual difficulty for undergraduate learners.

Empirical difficulty can be estimated from student performance using Classical Test Theory (CTT), which defines difficulty as the proportion of students answering correctly (p-value), and Item Response Theory (IRT), which models the interaction between latent ability and item characteristics [9,12,15]. Recent studies have explored the use of LLMs to approximate expert judgments of difficulty or to infer IRT parameters from item text [1,28]; however, findings suggest that surface textual features are weak predictors of latent difficulty compared to empirical response data.

This study evaluates whether LLM-generated difficulty estimates align with empirical undergraduate performance on basic arithmetic items. Beyond measuring alignment, we investigate a systematic source of misprediction: the distinction between curricular difficulty (what is expected to be easy based on instructional progression) and cognitive difficulty (what is actually difficult due to persistent misconceptions). We hypothesize that LLMs primarily encode curricular expectations, leading to systematic underestimation of items that are procedurally simple but conceptually challenging. Using responses from 770 Indonesian undergraduates and difficulty estimates from four LLM systems, we examine where and why these mismatches occur, and what they imply for the use of LLMs in educational assessment.

Research questions:

- **RQ** To what extent do LLM-based difficulty estimates align with empirical student performance on undergraduate basic arithmetic, and how does the disconnect between these measures reveal specific misconception-driven demands that models fail to capture?

## 2. RELATED WORKS

Item difficulty is traditionally measured using Classical Test Theory (CTT) and Item Response Theory (IRT). In the 2-parameter logistic (2PL) IRT model, item difficulty and discrimination parameters enable sample-invariant estimation under standard conditions and support fine-grained diagnostics [12,15]. Emerging work explores using large language models (LLMs) to approximate expert judgments of difficulty or infer psychometric properties directly from item text [23,28]. However, evidence suggests that surface textual features are weak predictors of latent difficulty compared to empirical response data, and many approaches rely on simulated responses or proxy indicators rather than real student performance [1].

LLMs are increasingly used to support assessment workflows, including item generation, grading, and feedback [4,8,37]. Recent studies have investigated whether LLMs can estimate item difficulty, with promising correlations reported in controlled settings [1,28]. However, these approaches largely depend on simulated learners or benchmark datasets, leaving open questions about how well LLM-generated difficulty reflects actual student performance, particularly in conceptually demanding domains.

Fractions are widely recognized as a challenging domain in mathematics education, with difficulties persisting into higher education [5,21]. Prior work shows that fraction understanding predicts later success in algebra and overall mathematical achievement [17,31]. These difficulties are often driven by persistent misconceptions, such as whole-number bias and incorrect reasoning about operations, which continue to affect even university students [10,11].

Despite this extensive literature, there is limited empirical data on fraction performance among undergraduates. This gap makes it difficult to assess whether LLM-generated difficulty estimates reflect young adult learner performance or primarily encode curriculum-based expectations.

In this study, we combine CTT and 2PL IRT with real response data from 770 Indonesian undergraduates to evaluate the alignment between LLM-generated difficulty estimates and empirical difficulty. By grounding the analysis in observed learner performance, we examine how discrepancies between model predictions and student outcomes reveal the limitations of LLM-based difficulty estimation.

## 3. METHOD

### 3.1 Participant

Participants were 770 second-year undergraduate students enrolled during 2020 to 2025 in mandatory mathematics courses at a large public university in Indonesia (79.74% female, 20.26% male; mean age = 19.2 years). All students had completed Indonesia-K13 mathematics curricula through Grade 9 – 12 (before the national shift to the Merdeka Curriculum in 2021). Participation was voluntary, and informed consent was obtained. The study used previously collected data approved for secondary analysis by the university ethics board (Authorization No. 511/In.23/L.1/TL.01/12/2025) and received additional approval from the Human Research Ethics Committee (Project ID: 50488).

### 3.2 Instrument

#### 3.2.1 *BMA Test*

The Basic Mathematics Ability (BMA) assessment consisted of 32 multiple-choice items covering arithmetic operations aligned with K13 Grades 3-7 standards. The distribution was as follows: whole number operations (n = 5), integer operations (n = 10), fraction operations (n = 11), decimal operations (n = 4), and exponents/roots (n = 2). Items were drawn from validated Indonesian textbooks (published by the Ministry of Education) and prior validated research [20]. Each item had four response options.

The present analysis focuses on eleven fraction items (Items 3, 11, 20, 23, 25, 27, 28, 29, 30, 31, 32) spanning addition, subtraction, multiplication, division, and magnitude comparison. These eleven items represent all fraction-related items from the 32-item assessment. The test was administered via Google Forms and paper-and-pencil formats in proctored classroom settings (90-minute time limit). Calculators were not permitted to ensure assessment of procedural fluency.

**Prompt for Generate Item Difficulty**

You are an expert in mathematics assessment design.

Evaluate the following item:

**ITEM**: [Full item picture with options]

**CONTEXT**:

- Target population: Indonesian undergraduate students
- All content taught in elementary school under Indonesian Kurikulum 2013
- Purpose: Evaluate item characteristics and estimate functional difficulty for undergraduate students on a Basic Mathematics Ability (BMA) instrument.
- Item Review: Identification of mathematical topic, domain, targeted skill, and curriculum grade alignment.
- Cognitive Demand: Assessment of skill difficulty, cognitive load, multi-step reasoning, and prerequisite concepts.
- Difficulty Scale: Constrained numerical scale reflecting automatic fluency, basic procedural skill, or basic multi-step reasoning.
- Provide an exact difficulty score (1 - 100 scale):
  1 - 10 = Automatic basic fluency
  11 - 20 = Basic procedural skill
  21 - 30 = Basic multi-step
  31+ = Higher cognitive load
- Justify your rating briefly.

**OUTPUT**: Structured item-level difficulty table with mathematical topic, domain, prerequisite concepts, multi-step reasoning, and curriculum alignment.

**Figure 1. Standardized Prompting Template That Used in All LLMs**

#### 3.2.2 *LLM Difficulty Estimation Prompt*

Difficulty predictions were obtained from four frontier LLM based AI systems:

1) Claude Sonnet 4.5 (Anthropic, September 2025)
2) Gemini 3 (Google DeepMind, November 2025)
3) ChatGPT 5.2 (OpenAI, GPT-5 series, December 2025)
4) Microsoft Copilot (Powered by GPT-5, December 2025)

For each item, each model was queried independently across four repetitions (N = 4 models × 5 repetitions × 32 items = **640 total difficulty scores**). A standardized prompting template was used as illustrated in Figure 1 (detailed outputs are available in the GitHub repository). We evaluated consistency across multiple runs using flip-rate, which measures the proportion of questions where the model produced different answers between runs [13].

## 3.3 Empirical Difficulty Measures

We calculated the empirical item difficulties using Classical Test Theory (CTT) and 2-parameter logistic (2PL) IRT. These approaches provide complementary perspectives on item difficulty. CTT offers intuitive, sample-dependent difficulty indices based on the proportion of correct responses, whereas IRT estimates latent item difficulty parameters (β) while accounting for differences in student ability levels. The 2PL model was selected instead of the 1PL (Rasch) model because preliminary inspection indicated substantial variation in item discrimination across the 32 problems (a range from 0.28 - 1.83, M = 1.20, SD = 0.37). Item spanned all five discrimination categories defined by Baker [2], from very low (n = 2) to very high (n = 2), with the majority falling in the moderate (n = 16) ranges. Although a detailed analysis of discrimination is beyond the scope of this study, this heterogeneity is inconsistent with the equal-discrimination assumption of the 1PL model and motives the use of 2PL. In the 2PL model, difficulty parameter can then be compared directly with LLMs-estimated difficulty [12].

We estimated 2PL IRT model parameters using marginal maximum likelihood (MML) via the *mirt* package in R [7]. IRT's item difficulty $\beta$ theoretically range from $-\infty$ to $+\infty$, but typically falls between $-3$ and $+3$ in practical testing. Values near 0 indicate average difficulty, negative values reflect easier items, and positive values correspond to harder items [2].

To evaluate alignment between LLM-generated difficulty estimates and empirical difficulty measures, we computed Spearman's rank correlation ($\rho$), Root Mean Squared Error (RMSE) and Mean Absolute Error (MAE). We rated consistency across multiple runs using two different metrics: (a) flip-rate, which measures the proportion of questions where the model produced different answers between runs [13], and (b) intra-class correlation (ICC) calculated using python *pingouin* package.

# 4. RESULTS

We distinguish between two forms of difficulty throughout the analysis: (1) curricular difficulty, reflected in LLM-generated estimates based on procedural and textual features, and (2) cognitive difficulty, reflected in empirical performance shaped by learner misconceptions. We evaluated the alignment between LLM-generated difficulty estimates and empirical difficulty using multiple metrics. Across models (32 arithmetic items), LLM predictions show moderate rank alignment with empirical difficulty (Spearman's ρ = 0.52 – 0.77; Table 1), indicating that models capture coarse ordering of item difficulty. Similar patterns were observed for both IRT and CTT measures.

However, rank alignment masks substantial discrepancies in absolute difficulty. RMSE and MAE values vary notably across models (Table 1), showing that models with similar correlations differ in their ability to estimate difficulty magnitude. This indicates that correlation alone overstates practical predictive accuracy.

**Table 1. Correlation Between Individual LLM Predictions and Empirical Difficulty Measures**

| LLM Model | IRT** | CTT** | RMSE (IRT) | MAE (IRT) | Rank |
|---|---|---|---|---|---|
| Claude Sonnet 4.5 | 0.700* | 0.776 | 0.809 | 0.638 | Best rank alignment |
| Gemini 3 Pro | 0.624* | 0.737 | 0.857 | 0.665 | Balanced |
| ChatGPT 5.2 | 0.567* | 0.716 | 0.974 | 0.721 | Worst magnitude |
| Microsoft Copilot | 0.515* | 0.595* | 0.943 | 0.727 | Best magnitude |
| CTT | 0.912* | | 0.366 | 0.257 | Empirical baseline |

*Significance in α = 0.05.
** Correlation using Spearman rho
For CTT, we are using (1 – p) to align the magnitude with other measurements

Figure 2 compares LLM-predicted difficulty with empirical IRT difficulty. For CTT, we are using (1 – p) to align the magnitude with other measurements. Non-fraction items follow the expected positive trend, indicating general alignment between model predictions and empirical measures. However, fraction items systematically deviate from this relationship. As empirical difficulty increases (β > 0), fraction items become much harder for students, but the increase in LLM-predicted difficulty is smaller than expected.

This judgement suggests that LLMs fail to capture sources of cognitive difficulty specific to fraction concepts. Although isolated cases of underestimation appear among non-fraction items, these do not form a consistent pattern. In contrast, the repeated and directional deviation observed for fraction items indicates a domain-specific bias rather than random error (also showed in flip rate at Figure 3). While within-model variance was observed, it did not consistently correspond to prediction error, suggesting that variability alone is not a reliable indicator of misalignment.

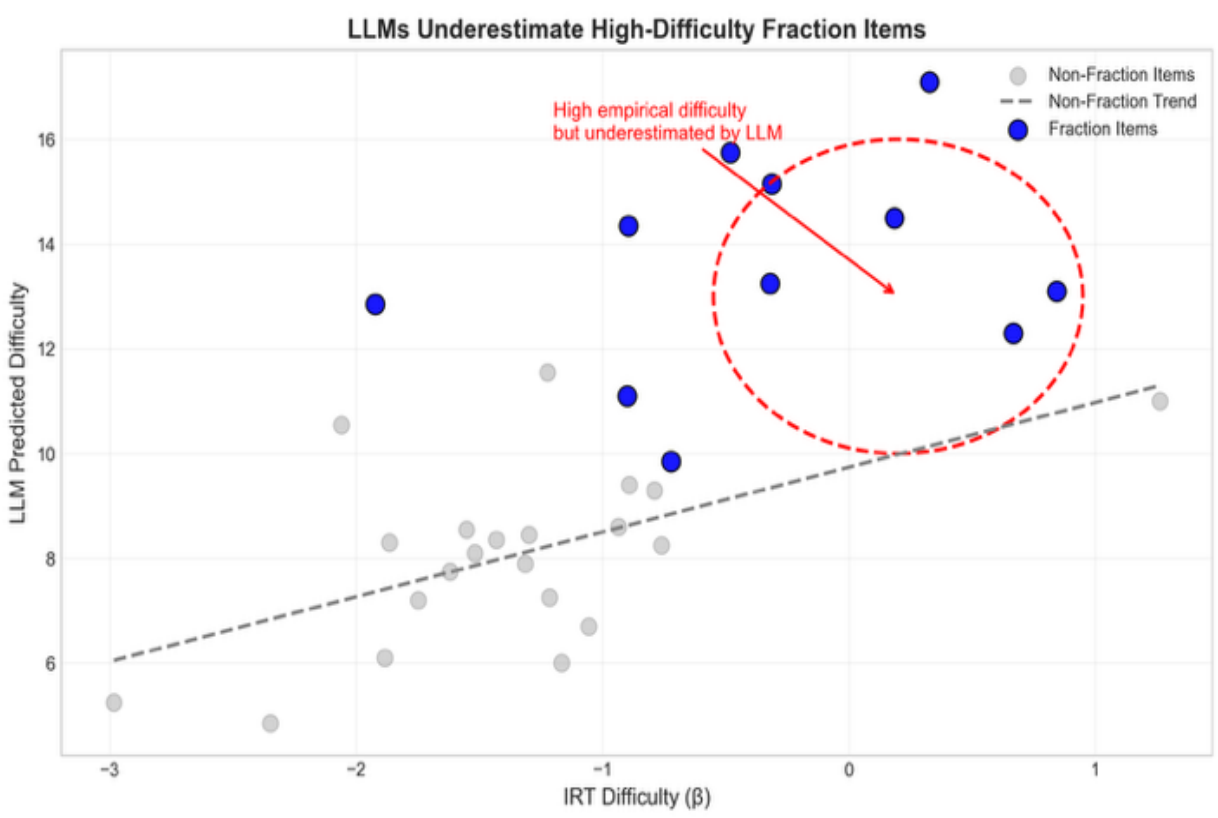

**Figure 2. As empirical difficulty increases, LLMs systematically assign lower-than-expected difficulty scores to fraction items.**

Fraction items account for the most severe misalignment between LLM-predicted and empirical difficulty. Table 2 lists the five most underestimated items, all of which were consistently rated as "easy" by LLMs despite empirical failure rates exceeding 66.8%. In the most extreme case, only 34.2% of students correctly solved division between whole number and fraction ($100 \div \frac{1}{2}$), although it was consistently categorized as a basic procedural task.

**Table 2. Fraction Items with the Largest LLM Underestimation of Empirical Difficulty**

| Item | LLM's Difficulty Estimation Range | | | | LLM Mean | IRT β | CTT |
|---|---|---|---|---|---|---|---|
| | Claude Sonnet | Gemini Pro | Chat GPT | Copilot | | | |
| P32 | 13 - 18 | 10 - 15 | 8 - 18 | 6 - 12 | 12.30 | 0.667 | 0.342 |
| P3 | 10 - 18 | 12 - 16 | 6 - 9 | 10 - 16 | 13.10 | 0.843 | 0.384 |
| P11 | 17 - 20 | 18 - 25 | 10 - 18 | 12 - 18 | 17.10 | 0.328 | 0.426 |
| P20 | 10 - 16 | 15 - 18 | 9 - 16 | 9 - 18 | 14.50 | 0.185 | 0.449 |
| P28 | 14 - 16 | 12 - 14 | 9 - 22 | 8 - 13 | 13.25 | -0.319 | 0.553 |

Note: LLM difficulty estimates represent the range across five runs for each model. LLM Mean is the average across all models and runs. IRT β is the 2PL difficulty parameter. CTT is the proportion of students answering correctly (p-value). Lower CTT values indicate higher empirical difficulty.

We refer to this systematic misclassification as the "Easy Trap": items that are procedurally simple but conceptually conflicting are consistently underestimated by LLMs.

Analysis of student response distributions (Table 3) reveals that errors are not random but reflect systematic misconception patterns. Common incorrect responses include "division makes smaller" beliefs, whole-number bias, inappropriate operation selection, and incorrect reasoning about magnitude. These patterns indicate that empirical difficulty is driven primarily by conceptual misunderstandings rather than procedural complexity.

**Table 3. Distribution of student responses and dominant misconception patterns for the five most underestimated fraction items**

| Item | Correct (%) | Common Incorrect Answer Pattern(s) | Misconception Type |
|---|---|---|---|
| **P32**: $100 \div \frac{1}{2}$ | 34.2% | 50, 100 | Operation misinterpretation ("division makes smaller") |
| **P3**: Estimate the sum of: $\frac{12}{13} + \frac{7}{8}$ | 38.4% | 19, 21 | Fraction magnitude misconception |
| **P11**: $2\frac{3}{5} + 1\frac{2}{3}$ | 42.6% | $3\frac{5}{8}$, $3\frac{8}{15}$ | Component-wise addition error |
| **P20**: $\frac{2}{3} - \frac{3}{5}$ | 44.9% | $\frac{1}{2}$, $\frac{1}{5}$ | Whole-number bias |
| **P28**: $\frac{12}{15} \div \frac{4}{15}$ | 55.3% | $\frac{3}{15}$, $\frac{1}{5}$ | Fraction division misconception |

Note: Correct (%) shows the proportion of students who answered correctly. Common Incorrect Pattern(s) lists the most frequent wrong answers chosen by students. Misconception Type describes the underlying conceptual error driving the incorrect responses.

This pattern is further supported by the flip rate analysis shown in Figure 3. Across most models, fraction items produced higher rates of inconsistent difficulty predictions than the overall item set (for 5 runs). Especially for Gemini, which shows almost 91% inconsistency during repetition, which means 10 out of 11 fractions get different difficulty estimation more than 3 times. This suggests that LLM judgments on misconception-driven fraction problems are not only systematically misaligned with empirical student difficulty, but also less stable across repeated evaluations. The consistently higher flip rates for fraction items indicate that conceptually conflicting problems introduce greater uncertainty in LLM difficulty estimation compared to arithmetic items overall.

Taken together, these results suggest that LLM-generated difficulty estimates align more closely with procedural complexity and curricular expectations, but fail to fully capture cognitive difficulty driven by persistent misconceptions.

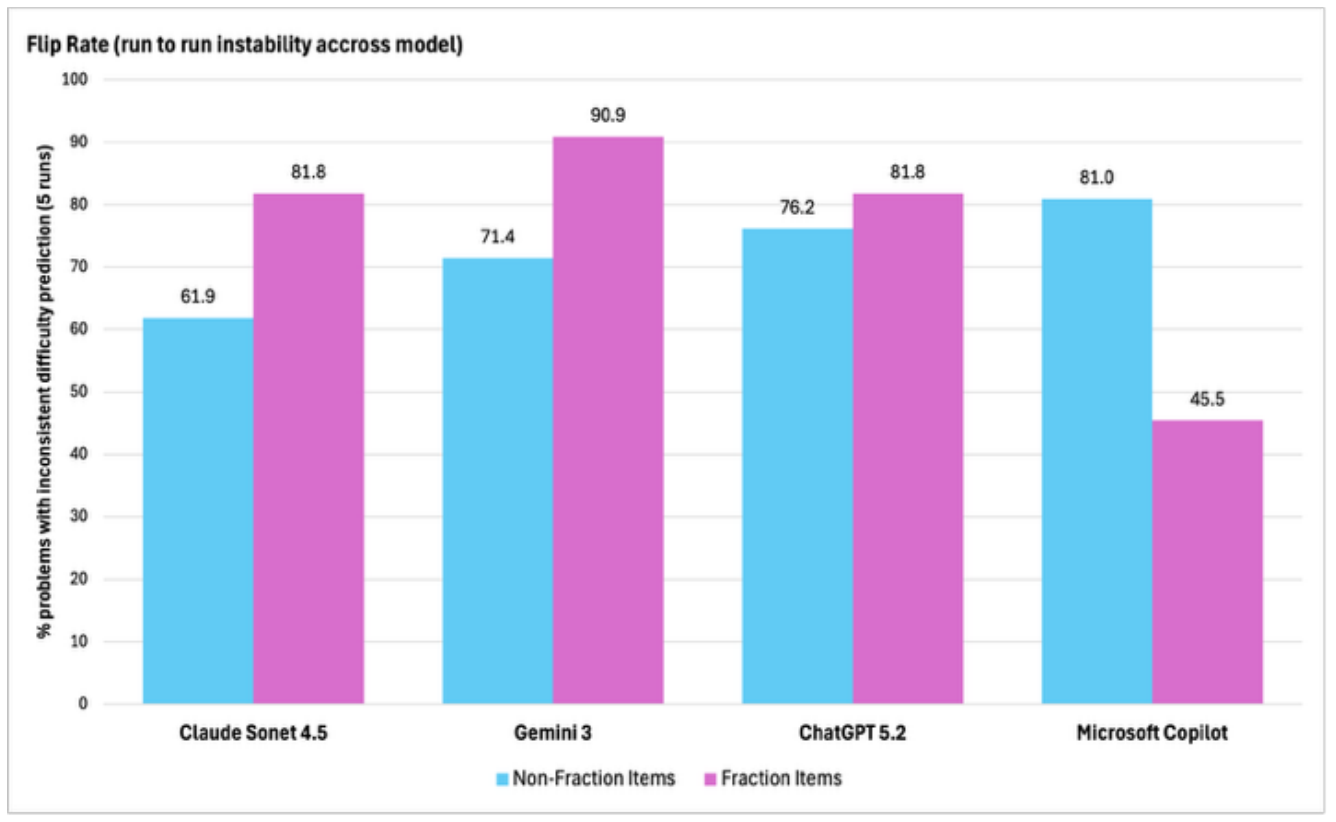

**Figure 3. Flip Rate (Run-to-run Instability Across Models)**

## 5. DISCUSSION

Our findings reveal a systematic distinction between curricular and cognitive difficulty in LLM-based estimation. While LLMs achieve moderate rank alignment with empirical measures, they consistently underestimate items whose difficulty is driven by persistent misconceptions. This suggests that LLMs rely primarily on surface-level procedural cues and curricular sequencing, rather than modeling the cognitive processes that shape student performance. As a result, LLM-generated difficulty estimates reflect what should be easy according to instruction, rather than what is actually difficult for learners.

Error analysis (Table 3) provides insight into the cognitive sources of this misalignment. Undergraduates continue to retain foundational misunderstandings typically associated with earlier grades. Common errors—"division makes smaller" reasoning (Item P32), inappropriate operation selection (Item P20), and whole-number overgeneralization (Items P11, P28)—indicate many students lack robust mental models of fractions. These patterns align with established research on fraction difficulty, particularly whole-number bias and magnitude understanding [5,17,18,24,32]. Our study extends this line of work with a higher-education perspective: despite completing 12 years of schooling, many students still lack stable conceptual understanding and frameworks for elementary fraction operations.

Stability analyses reveal that fraction items are vulnerable to run-to-run variability. Figure 3 show high flip-rates meaning that users may receive different difficulty labels for identical items across attempts. This align with argument by Gonzales [13] that GPT-4 family models change their answer on ~35% of problems in 3 repetitions. This volatility reflects both stochastic generation and inconsistent sensitivity to cognitive features, underscoring the need for response aggregation and uncertainty reporting when difficulty scores inform instructional decisions or assessment design.

These results have direct implications for educators and institutions increasingly relying on LLM-generated items as part of classroom assessment. In Indonesian higher education, classroom practice already uses LLM-generated items to evaluate student understanding [25,26,34]. Our findings caution against depending on single difficulty estimates for high-stakes decisions, as our repeated query analyses show that LLMs often produce varying interpretations of identical prompts across users and attempts—instability also documented by Gonzales [13] and Castleman et al. [6] across leading LLMs on mathematics problems. Instead, LLM estimates should be treated as provisional heuristics requiring empirical validation, not ground validated difficulty values. The misalignment between algorithmic and cognitive difficulty means LLMs may systematically misclassify which items will challenge students.

This difficulty misalignment is particularly important for educational data mining applications. In adaptive testing and automated item generation, difficulty estimates play a central role in sequencing content and personalizing learning experiences. When misconception-driven items are systematically rated as easier than they actually are, assessment systems may select inappropriate items, miscalibrate learner ability, and reduce the effectiveness of instructional interventions. These findings suggest that LLM-based assessment systems should be grounded in empirical learner data rather than relying solely on model-generated difficulty estimates. Overall, this study should be interpreted as an exploratory baseline for understanding how LLMs estimate mathematical difficulty in misconception-driven contexts. These results provide an initial foundation for future work on cognitively informed LLM-based assessment systems.

Several limitations should be considered when interpreting these findings. First, this study focuses on fraction arithmetic among Indonesian undergraduate students. Although fractions are well known for misconception-driven errors [3,17], it remains unclear whether the "Easy Trap" phenomenon also appears in other mathematical domains on undergraduate learning such as algebra, geometry, or calculus. Future studies should examine whether similar patterns emerge across different topics. Second, while the dataset included 770 students, the analysis was based on only 32 arithmetic items, with the clearest evidence of systematic underestimation found in 11 fraction items. A broader evaluation involving more items and mathematical domains would provide stronger evidence regarding the consistency of LLM difficulty misalignment. In addition, the data were collected from a single institution, so further validation across institutions and cultural contexts is needed. Third, our prompting approach was intentionally designed to reflect realistic educational use, where educators ask LLMs to estimate item difficulty without detailed information about student misconceptions or error patterns. Although this mirrors current practice in AI-assisted assessment design, richer prompts that include misconception patterns or cognitive analysis may improve alignment between LLM predictions and empirical difficulty. Future research should investigate whether misconception-informed prompting can reduce this systematic bias.

# 6. CONCLUSION

This study shows that LLM-based difficulty estimation captures the relative ordering of item difficulty but systematically fails to represent cognitive difficulty driven by misconceptions. As a result, fraction items—where performance depends on conceptual understanding rather than procedural complexity—are consistently underestimated. We interpret this as evidence that LLMs encode curricular expectations rather than learner-centered difficulty, leading to a measurable and predictable bias.

Taken together, these findings suggest that while LLMs are promising tools for rapid item prototyping and scalable content generation, their effective use in assessment requires approaches that integrate model efficiency with empirical validation, psychometric calibration, and human expertise. Although this study focuses on fraction arithmetic, this domain provides a well-established testbed for examining persistent misconceptions and cognitive bottlenecks in mathematics learning. The observed misalignment therefore points to broader limitations of LLM-based assessment tools when applied to conceptually demanding domains.

# 7. ACKNOWLEDGMENTS

This work was supported by the Indonesia Endowment Fund for Education (LPDP), Ministry of Finance of the Republic of Indonesia.